\documentclass{article}

\usepackage{amssymb,amsmath,graphicx}
\begin{document}

\title{Stability of Einstein-Aether Cosmological Models}

\author{\sc
P. Sandin$^{1,2}$\thanks{Electronic address:
{\tt patrik.sandin@aei.mpg.de}}\ ,
B. Alhulaimi$^{2}$%\thanks{Electronic address: {\tt xxx.xxx@mathstat.dal.ca}}
,\ and
A. Coley$^{2}$\thanks{Electronic address: {\tt aac@mathstat.dal.ca}}%\  and
%M. Stevens$^{2}$%\thanks{Electronic address: {\tt xxx.xxx@mathstat.dal.ca}}
\\
$^{1}${\small\em Max-Planck-Institut f{\"u}r Gravitationsphysik (Albert-Einstein-Institut)}\\
{\small\em  Am M{\"u}hlenberg 1, D-14476 Potsdam, Germany}\\
$^{2}${\small\em Dept. of Mathematics, Dalhousie University, B3H 3J5 Halifax, Canada}
}

\maketitle

\begin{abstract}
We use a dynamical systems analysis to investigate the future behaviour of
Einstein-Aether cosmological models with a scalar field coupling to the expansion of the aether and a non-interacting perfect fluid. The stability of the equilibrium solutions are analysed and the results are compared with the standard inflationary cosmological solutions and previously studied cosmological Einstein-Aether models.
\end{abstract}
%PACS numbers 04.50.Kd, 04.20.Fy, 98.80.Cq

\section{Introduction}

Einstein-Aether theory \cite{DJ, Barrow} consists of general relativity coupled, at second derivative order, to a dynamical timelike unit vector field -- the aether. It is one of several proposed models of early universe cosmology which incorporate a violation of Lorentz invariance \cite{eit}. In this effective field theory approach, the aether vector field, $u_{a}$, and the metric tensor $g_{ab}$ together determine the local spacetime structure.

We shall discuss the late time dynamics of Einstein-Aether cosmological models; in particular we explore the impact of Lorentz violation on the inflationary scenario \cite{infl}, which provides one of the simplest ways to describe various aspects of the physics of the early universe in standard cosmology. More precisely, we study the inflationary scenario in the scalar-vector-tensor theory where the vector is constrained to be unit and time like, and investigate whether an example of the large class of the inflationary solutions proposed \cite{Barrow} are stable when spatial curvature perturbations are considered.

\subsection{Einstein-Aether Cosmology}

In an isotropic and spatially homogeneous Friedmann
universe with expansion scale factor $a(t)$ and comoving proper time $t$,
the aether field will be aligned with the cosmic frame and is related to the
expansion rate of the universe.
The Einstein equations are generalized by the contribution of an additional
stress tensor for the aether field. If the universe contains a single
self-interacting scalar field $\phi$ (e.g., a scalar inflation which would
dominate in any inflationary epoch), with a self interaction potential $V$
that can now be a function of $\phi $ and the expansion rate
$\theta =3\dot{a}/a=3H,$ then the modified stress tensor is
 \cite{DJ}

\begin{equation}
T_{ab}=\nabla _{a}\phi \nabla _{b}\phi -(\frac{1}{2}\nabla _{c}\phi \nabla
^{c}\phi -V+\theta V_{\theta })g_{ab} + \dot{V_{\theta}}(u_a u_b - g_{ab}). \label{tab}
\end{equation}%

This corresponds to an effective fluid with pressure $p$ and density $\rho $
of the form
$\rho =\frac{1}{2}\dot{\phi}^{2}+V-\theta V_{\theta }$ and
$p=\frac{1}{2}\dot{\phi}^{2}-V+\theta V_{\theta }+\dot{V}_{\theta }$,
where $V(\phi ,\theta )$.

The energy-momentum conservation law or Klein-Gordon eqn. is
\begin{equation}
\ddot{\phi}+\theta\dot{\phi}+V_{\phi }=0,  \label{1}
\end{equation}%
The  augmented Friedmann equation (with the addition  of the aether
stress to the energy density, and where $8\pi G=1=c$, and $G$ is the renormalized
gravitational constant) is given by:

\begin{equation}
\frac{1}{3}\theta^{2}=\rho  + \frac{1}{2}\dot{\phi}^{2}+V-\theta V_{\theta }-\frac{k}{a^{2}},
\label{2}
\end{equation}%
where $k$ is the curvature parameter, and the Friedmann metric is given by
\begin{equation}
ds^{2}=dt^{2}-a^{2}(t)\{\frac{dr^{2}}{1-kr^{2}}+r^{2}d\vartheta
^{2}+r^{2}\sin ^{2}\vartheta d\varphi ^{2}\}.
\label{feq}
\end{equation}
[In \cite{Barrow} $k$ was set equal to zero].  The Raychaudhuri eqn. follows
from the differentiation of the Friedmann eqn.

\subsection{Exponential Potentials}

Exponential potentials of the form $V_{0} e^{-\lambda \phi}$ arise naturally in various higher dimensional frameworks, such as in Kaluza-Klein theories and supergravity \cite{super},
and the dynamical properties of the positive exponential potentials leading to
inflation in the Friedmann-Robertson-Walker (FRW) model have been widely
studied \cite{lucc,dyn,coleybook}. In Einstein-Aether theories there might exist a coupling between the scalar field and the aether field through the aether expansion scalar $\theta$, which requires one to consider more general types of potentials than the standard exponential form.
In \cite{Barrow} Barrow proposes an ansatz of the form

\begin{equation}
V(\theta ,\phi )=V_{0}\exp [-\lambda \phi ]+\sum_{r=0}^{n}a_{r}\theta
^{r}\exp [(r-2)\lambda \phi /2],  \label{3}
\end{equation}%
where $V_{0},\lambda $ and $\{a_{r}\}$ are constants, and concludes that there exist power law solutions of the Einstein-Aether-generalized combined Friedmann-Klein Gordon system on the form
\begin{align}
\phi &= \frac{2}{\lambda} \ln t, \\
a &= t^B, \\
\theta &= 3 \frac{\dot{a}}{a} = 3Bt^{-1},
\end{align}
where $B$ is a solution of the polynomial equation
\begin{equation}\label{B_eqn}
B + \frac{1}{2} \sum_{r = 0}^{n}ra_r(3B)^{r - 1} = \frac{2}{\lambda}.
\end{equation}
He presents two potentials for which explicit solutions to \eqref{B_eqn} can be found, namely potentials where $a_i = 0\ \forall i$ and potentials where only $a_2$ is different from zero. This corresponds to, respectively, the ordinary exponential potential and an exponential potential plus a term quadratic in the expansion. We shall consider one of Barrow's examples where the exponential part couples to the expansion scalar.

\section{The Model}

We shall study the case:
\begin{equation}
V(\theta ,\phi )=V_{0} e^{-\lambda \phi} +
a_{1}  \sqrt{V_{0} }\theta
e^{- \frac{1}{2} \lambda \phi}
+ a_{2} \theta^2,  \label{33}
\end{equation}
where, for convenience, we have renormalized the constant $a_1$ (the constant $a_2$ can be absorbed \cite{DJ} and will not play an essential role in the dynamical analysis). The potential $V(\theta ,\phi )$ can be assumed to be positive definite, but this does not imply that the constants $a_{1}, a_{2}$ necessarily are positive, as can be seen in the positive definite potential $(e^{- \frac{1}{2} \lambda \phi}- \frac{1}{2} \theta)^2$ where $a_1$ is negative and $a_2 = {a_1}^2/4$.

If $a_{1}, a_{2}$ are small, the potential can be
thought of as a perturbation of the standard exponential potential. For very large constants $a_{1}, a_{2}$ we can study non-perturbative generalizations.

For the potential \eqref{33} the augmented Friedmann equation \eqref{2} becomes
\begin{equation}
1 = \frac{3}{2(1+ 3a_2)}\left( \frac{\dot{\phi}}{\theta} \right)^2 + \frac{3 V_0}{1 + 3a_2} \frac{e^{-\lambda\phi}}{\theta^2} - \frac{3k}{(1 + 3a_2)} \frac{1}{a^2 \theta^2}
\end{equation}
where we have normalized the equation with a factor proportional to the square of the expansion. The normalized Friedmann equation suggest a suitable set of expansion normalized variables:
\begin{equation}\label{var_def}
\Psi := \sqrt{\frac{3}{2(1 + 3a_2)}}\frac{\dot{\phi}}{\theta}, \quad \Phi : = \sqrt{\frac{3V_0}{(1 + 3a_2)}}\frac{e^{-\lambda \phi /2}}{\theta}, \quad K:= \frac{3k}{(1 + 3a_2)} \frac{1}{a^2 \theta^2},
\end{equation}
assuming that $V_0$ is a positive constant and that $a_2$ is larger than $-1/3$. In terms of these variables the Friedmann equation assumes the simple form
\begin{equation}\label{Friedmann}
1 = \Psi^2 + \Phi^2 - K,
\end{equation}
and the Raychaudhuri equation (expressed in terms of the deceleration parameter $q$) becomes
\begin{equation}
q:= -3(\frac{\dot{\theta}}{\theta^2} + \frac{1}{3}) = 2\Psi^2 - \Phi^2  -\frac{3\lambda a_1}{2\sqrt{2}} \Psi \Phi,
\end{equation}
Using the Raychaudhuri equation one can express the Klein-Gordon equation as a first
order ordinary differential equation completely in terms of the expansion normalized
variables, and an expansion normalized time: $\frac{d\tau}{d t} = 3\theta^{-1}$:
\begin{equation}\label{dPsi}
\frac{d\Psi}{d \tau} = - (2 - 2\Psi^2 + \Phi^2)\Psi + \bar{\lambda} \Phi^2 + \bar{a} (1 - \Psi^2)\Phi
\end{equation}
where the constants $\bar{a}$ and $\bar{\lambda}$ are defined through
\[
\bar{a} = \frac{3\lambda a_1}{2\sqrt{2}}, \quad \bar{\lambda} = \lambda \sqrt{\frac{3(1+3a_2)}{2}}.
\]
The evolution equation for $\Phi$ is directly given from the definition of $\Phi$ and the Klein-Gordon and Raychaudhuri equations:
\begin{equation}\label{dPhi}
\frac{d\Phi}{d \tau}  = (1 + 2\Psi^2 - \Phi^2 - \bar{a}\Psi\Phi - \bar{\lambda}\Psi)\Phi.
\end{equation}

The equations \eqref{dPsi},\eqref{dPhi} constitute an autonomous system
of first order differential equations. The invariant set $\Phi=0$ corresponds to a model with a free scalar field, and it divides the two-dimensional (2D) state space into a region with an expanding universe ($\Phi \ge 0$) and one corresponding to a contracting universe ($\Phi \le 0$).

The curvature $K$ is determined from the Friedmann equation \eqref{Friedmann}, but an auxiliary evolution equation can be derived from \eqref{dPsi},\eqref{dPhi}: ${d K}/{d \tau} = [4\Psi^2 - 2\Phi^2 -2\bar{a}\Psi\Phi] K$, showing that $K=0$ is an invariant set that also partitions the state space into two disjoint regions: a bounded negative curvature region $(\Psi^2 + \Phi^2 < 1)$, and an unbounded positive curvature region $(\Psi^2 + \Phi^2 > 1)$.

Note that $a_2$ does not explicitly appear in the equations; it corresponds to a term in the potential proportional to the square of the expansion and can be absorbed by a rescaling of the expansion scalar, but the expansion normalized system is invariant under such rescalings.

Defining $ |\bar{A}|= \sqrt{(9 - \bar{\lambda}^2 + \bar{a}^2)}$,
the equilibrium points ($p_i$) of the system are given by

\begin{align*}
& \text{Equilibria} & & & & & & K,\ q \\
\hline
\\
& p_1:\ & & \Psi = 0, & & \Phi = 0, & & -1,\ 0\\
& p_2:\ & & \Psi = \pm 1, & & \Phi = 0, & & 0,\ 2\\
& p_3:\ & & \Psi = \frac{3\bar{\lambda} + |\bar{a}\bar{A}|}{9 + \bar{a}^2}, & & \Phi = \frac{-\bar{\lambda} \bar{a}^2 + 3\, |\bar{a}\bar{A}|}{(9 + \bar{a}^2)\bar{a}}, & & 0,\ \frac{\bar{\lambda}\bar{a}|\bar{A}| - 9 + 3\bar{\lambda}^2 - \bar{a}^2}{( 9 +\bar{a}^2)}\\
& p_4:\ & & \Psi = \frac{3\bar{\lambda} - |\bar{a}\bar{A}|}{9 + \bar{a}^2}, & & \Phi = -\frac{\bar{\lambda} \bar{a}^2 + 3\, |\bar{a}\bar{A}|}{(9 + \bar{a}^2)\bar{a}}, & & 0,\ -\frac{\bar{\lambda}\bar{a}|\bar{A}| + 9 - 3\bar{\lambda}^2 + \bar{a}^2}{( 9 +\bar{a}^2)}\\
& p_5:\ & & \Psi = \frac{1}{\bar{\lambda}}, & & \Phi = \frac{-\bar{a} + \sqrt{\bar{a}^2+8}}{2\bar{\lambda}}, & & -\frac{2\bar{\lambda}^2 - 6 - \bar{a}^2 + \bar{a}\sqrt{\bar{a}^2 + 8}}{2\bar{\lambda}^2},\ 0\\
& p_6:\ & & \Psi = \frac{1}{\bar{\lambda}}, & & \Phi = \frac{-\bar{a} - \sqrt{\bar{a}^2+8}}{2\bar{\lambda}}, & & \frac{-2\bar{\lambda}^2 + 6 + \bar{a}^2 + \bar{a}\sqrt{\bar{a}^2 + 8}}{2\bar{\lambda}^2},\ 0\\
\\
\hline
\end{align*}

The stationary solution $p_6$ has $\Phi < 0$ for all values of
$\bar{a} \in \mathbb{R}$, $\bar{\lambda} \in \mathbb{R}^+$, and
corresponds to a contracting universe. We shall in the following only consider expanding solutions with vanishing or negative curvature and will therefore ignore this point since it lies outside the region of interest $K \leq 0,\ \Phi \geq 0$. Points $p_1$ and $p_2$ always satisfies these conditions. $p_1$ is always a saddle, $p_2^+$ saddle or source, and $p_2^-$ always a source. We are most interested in the equilibra $p_3$, $p_4$, and $p_5$. The points $p_3$, $p_4$, and $p_5$ are contained in this region only for a restricted, partially overlapping, range of values in $(\bar{\lambda},\ \bar{a})$-space.

\subsubsection*{Equilibrium Point $p_3$:}

Range of validity:
\begin{equation}
\Phi_{p_3} \geq 0\ \text{when}\ (\bar{\lambda} \leq 3,\ \bar{a} \geq 0),\ \text{or}\ (\bar{\lambda} \geq 3,\ \bar{a} \leq - \sqrt{\bar{\lambda}^2 - 9}).
\end{equation}

Eigenvalues:

\begin{align*}
\mu_\pm  &= \frac{-5|\bar{A}|^2 + 4\bar{\lambda}^2 + 3\bar{\lambda}|\bar{a}\bar{A}| \pm \sqrt{\mathcal{B} + \mathcal{C}}}{2 (9 + \bar{a}^2)} % \\
% \mu_2  &= \frac{-5|\bar{A}|^2 + 4\bar{\lambda}^2 + 3\bar{\lambda}|\bar{a}\bar{A}| - \sqrt{\mathcal{B} + \mathcal{C}}}{2 (9 + \bar{a}^2)}
,
\end{align*}
where
\[
\mathcal{B} = \bar{a}^4 + 18\bar{a}^2 + 15\bar{a}^2\bar{\lambda}^2 + 81 + 54\bar{\lambda}^2 - \bar{\lambda}^4\bar{a}^2 + \bar{\lambda}^2\bar{a}^4 + 9\bar{\lambda}^4, \quad \mathcal{C} = 2\bar{\lambda}(3\bar{\lambda}^2 + \bar{a}^2 + 9)|\bar{a}\bar{A}|.
\]

Discussion: The point $p_3$ is a sink and inflationary when $0 < \bar{a}$ and $\bar{\lambda}^2 < \frac{1}{2} ( \bar{a}^2 + 6 - \bar{a}\sqrt{\bar{a}^2 + 8})$.

\subsubsection*{Equilibrium Point $p_4$:}

Range of validity:
\begin{equation}
\Phi_{p_4} \geq 0\ \text{when}\ (\bar{a} \leq 0,\ \bar{a}^2 \geq \bar{\lambda}^2 - 9).
\end{equation}

Eigenvalues:

\begin{align*}
\mu_\pm  &= \frac{-5|\bar{A}|^2 + 4\bar{\lambda}^2 + 3\bar{\lambda}|\bar{a}\bar{A}| \pm \sqrt{\mathcal{B} - \mathcal{C}}}{2 (9 + \bar{a}^2)}%  \\
% \mu_2  &= \frac{-5|\bar{A}|^2 + 4\bar{\lambda}^2 + 3\bar{\lambda}|\bar{a}\bar{A}| - \sqrt{\mathcal{B} - \mathcal{C}}}{2 (9 + \bar{a}^2)}
.
\end{align*}

Discussion:  The point $p_4$ is a sink when $\bar{a} < 0$ and $\bar{\lambda}^2 < \frac{1}{2} ( \bar{a}^2 + 6 - \bar{a}\sqrt{\bar{a}^2 + 8})$. It is inflationary only for the subset of this region where $\bar{\lambda}^2 < \frac{1}{2} ( \bar{a}^2 + 6 + \bar{a}\sqrt{\bar{a}^2 + 8})$.

\subsubsection*{Equilibrium Point $p_5$:}

Range of validity:
\begin{equation}
\Phi_{p_5} \geq 0\ \text{always, but}\ K_{p_5} \leq 0\ \text{only when}\ 2\bar{\lambda}^2 \geq 6 + \bar{a}^2 - \bar{a}\sqrt{\bar{a}^2 + 8}.
\end{equation}

Eigenvalues:

\begin{align*}
\mu_\pm  &= -1 \pm \tfrac{1}{2\bar{\lambda}}\sqrt{48 + 22\bar{a}^2 - 12\bar{\lambda}^2 + 2\bar{a}^4 - 2\bar{a}^2\bar{\lambda}^2 - 2\bar{a}(7 + \bar{a}^2 - \bar{\lambda}^2)\sqrt{\bar{a}^2 + 8}}% ,  \\
% \mu_2  &=  -1 - \tfrac{1}{2\bar{\lambda}}\sqrt{48 + 22\bar{a}^2 - 12\bar{\lambda}^2 + 2\bar{a}^4 - 2\bar{a}^2\bar{\lambda}^2 - 2\bar{a}(7 + \bar{a}^2 - \bar{\lambda}^2)\sqrt{\bar{a}^2 + 8}}
.
\end{align*}

Discussion: The point $p_5$ is a sink when $\bar{\lambda}^2 > \frac{1}{2} ( \bar{a}^2 + 6 - \bar{a}\sqrt{\bar{a}^2 + 8})$.

\begin{figure}[h!]
\includegraphics[width=\textwidth]{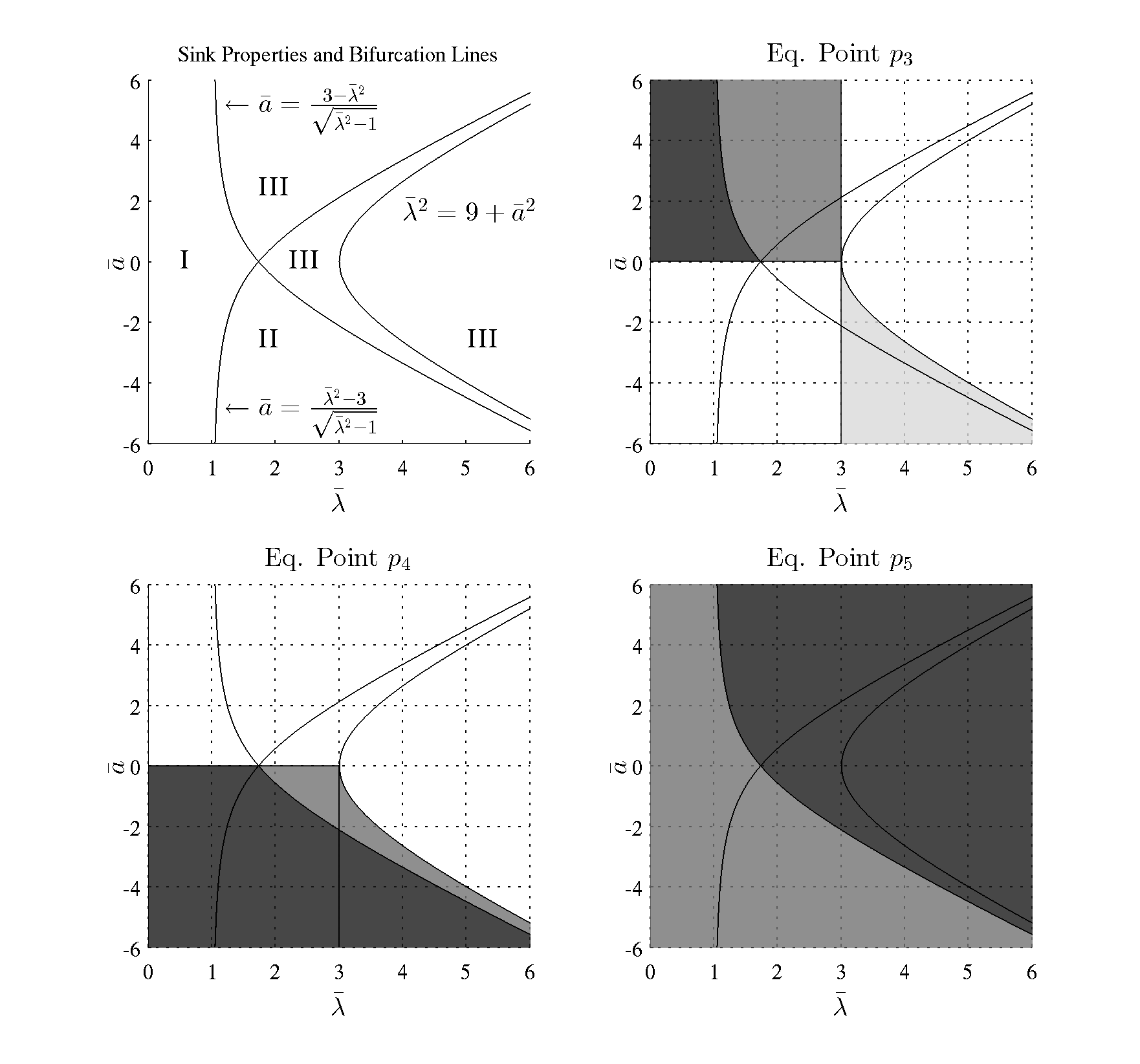}\caption{The first figure shows the properties of the local sink and the curves where bifurcations occurs in parameter space. Region I corresponds to a spatially flat inflationary universe, region II to a flat decelerating universe, and region III to a negatively curved universe with a constant expansion rate. The next three pictures show the sink (dark)/saddle (medium)/ source (light) properties of the individual equilibrium points (only $p_3$ is ever a source here).}
\label{sinkplots}
\end{figure}

Figure \ref{sinkplots} shows the relevant properties of the future attractor of the system and the stability properties of the equilibrium pints $p_3$, $p_4$, $p_5$.

\newpage

\section{The Model with Matter:}

If, in addition to the scalar field, there also exists a barotropic perfect fluid with linear equation of state, $p=(\gamma-1)\rho$ (satisfying $\frac{2}{3} < \gamma < 2$), we will get extra terms in the equations corresponding to the fluid energy density. Assuming that there is no transfer of energy between the scalar field and the fluid except through gravitation, we get an additional evolution equation for the fluid energy density coming from the matter energy conservation equation. A normalized energy density is defined through

\[\Omega := \frac{3\rho}{(1 + 3a_2)\theta^2}.\]

The augmented Friedmann and Raychaudhuri equations become

\begin{align}\label{Friedmann22}
1 &= \Omega+\Psi^2 + \Phi^2 - K, \\
q &:= -3(\frac{\dot{\theta}}{\theta^2} + \frac{1}{3}) =
\frac{1}{2}(3{\gamma}-{2})\Omega + 2\Psi^2 -\Phi^2 -{\bar{a}} \Psi \Phi,
\end{align}
and the evolution equations become:

\begin{align}\label{dPsi2}
\frac{d\Psi}{d \tau}
&= - (2 - 2\Psi^2 + \Phi^2-\frac{1}{2}({3\gamma}-2)\Omega)\Psi + \bar{\lambda} \Phi^2 + \bar{a} (1 - \Psi^2)\Phi, \\
\label{dPhi2}
\frac{d\Phi}{d \tau}  &= (1 + 2\Psi^2 - \Phi^2 - \bar{a}\Psi\Phi -
\bar{\lambda}\Psi+\frac{1}{2}({3\gamma}-2)\Omega)\Phi, \\
\label{dOmega}
\frac{d\Omega}{d \tau} &= (-(3\gamma-2)(1-\Omega) +4\Psi^2 -2\Phi^2 -2 \bar{a}\Psi\Phi)\Omega.
\end{align}

Note that $\Omega = 0$ defines an invariant set of the three-dimensional
autonomous system
of first order differential equations. The equilibrium points $p_i$ with  $\Omega = 0$ from the previous section are also equilibrium points of the extended system \eqref{dPsi2}-\eqref{dOmega}. Linearization about the points will result in the same eigenvalues as before for the eigendirections contained in the $\Omega=0$ subspace, but will also pick up a new eigenvalue from the influence of the fluid on the dynamics. The additional eigenvalues are:

\begin{align*}
&p_1:\ & & \mu_3 = -(3\gamma-2)\\
&p_2(\pm):\ & & \mu_3 = -3(\gamma-2)\\
& p_3:\ & & \mu_3 = -\frac{3\gamma \bar{a}^{2}+ 27\gamma -6\bar{\lambda}^{2}-2\bar{\lambda}\sqrt{\bar{a}^{2}(\bar{a}^2+9-\bar{\lambda}^{2})}}{9+\bar{a}^{2}}\\
& p_4:\ & & \mu_3 = -\frac{3\gamma \bar{a}^{2}+ 27\gamma -6\bar{\lambda}^{2}+2\bar{\lambda}\sqrt{\bar{a}^{2}(\bar{a}^2+9-\bar{\lambda}^{2})}}{9+\bar{a}^{2}}\\
& p_5:& & \mu_3 = -(3\gamma-2) & & % \\
% &p_6:& & \mu_3 = -(3\gamma-2) & & \\
\end{align*}

Equilibrium points $p_1$, $p_2$, and $p_5$ all gain an eigenvalue that is negative for all $\gamma > \frac{2}{3}$. The new eigenvalue transforms the source(s) to saddle(s) but retains $p_5$ as a sink in the same range as before. How it affects the points $p_3$ and $p_4$ is not immediately obvious, but as we shall see below it does not affect the properties of the sinks when $\gamma$ is in the range $(\frac{2}{3},2)$.

\subsubsection*{Equilibrium Point $p_3$:}

The point remains a sink if, in addition to the conditions in the previous section, also
\begin{eqnarray}\label{p3mattersinkcond}
2\bar{\lambda}^{2} + \frac{2}{3}\bar{\lambda}|\bar{a}\bar{A}| < \gamma (\bar{a}^{2}+9).
\end{eqnarray}
Solving for $\bar{\lambda}^2$ we get
\[ 2\bar{\lambda}^2 < \bar{a}^2 + 9\gamma - |\bar{a}|\sqrt{\bar{a}^2 + 18\gamma - 9\gamma^2}%, \quad \text{or} \quad 2\bar{\lambda}^2 > \bar{a}^2 + 9\gamma + |\bar{a}|\sqrt{\bar{a}^2 + 18\gamma - 9\gamma^2}
.\]
% Only the first inequality have the possibility to be compatible with the conditions in the previous section for the point to remain a sink.
For $\gamma = \frac{2}{3}$ this condition reduces to
\[\bar{\lambda}^2 < \frac{1}{2} ( \bar{a}^2 + 6 - \bar{a}\sqrt{\bar{a}^2 + 8}),\]
which is the same restriction as before. Since the r.h.s. of \eqref{p3mattersinkcond} is increasing with $\gamma$ we have that $p_3$ is a sink in the same range as before for all $\gamma \geq \frac{2}{3}$.

% For example, for the special case when  $\gamma=1,\bar{\lambda}=1$ and for any value of  $\bar{a}$, the equation \eqref{p3mattersinkcond}  becomes:

% \begin{equation}
% 5\bar{a}^4+94\bar{a}^2+441 > 0
% \end{equation}
% which is always true for any value of $\bar{a}$ (and thus the conditions on $\bar{a}$ in this case are the same as in the matter free case).

\subsubsection*{Equilibrium Point $p_4$:}
The third eigenvalue is negative if
\begin{eqnarray}\label{p4mattersinkcond}
2\bar{\lambda}^{2} - \frac{2}{3}\bar{\lambda}|\bar{a}\bar{A}| < \gamma (\bar{a}^{2}+9),
\end{eqnarray}
which implies
\[ 2\bar{\lambda}^2 < \bar{a}^2 + 9\gamma + |\bar{a}|\sqrt{\bar{a}^2 + 18\gamma - 9\gamma^2}.
\]
For $\gamma =  \frac{2}{3}$ this inequality also coincides with with one of the previous restrictions on $\bar{\lambda}$ required for $p_4$ to be a sink. Larger $\gamma$ gives a weaker restriction and hence is $p_4$ a sink in the same range as before if we have $\gamma \geq \frac{2}{3}$.

% \subsubsection*{Equilibrium Point $p_5$:}

% % Range of validity:
% % \begin{equation}
% % \Phi_{\bar{p_5}} \geq 0\ \text{always, but}\ K_{\bar{p_5}} \leq 0\ \text{only when}\ 2\bar{\lambda}^2 \geq 6 + \bar{a}^2 - \bar{a}\sqrt{\bar{a}^2 + 8}.
% % \end{equation}

% The point $p_5$ is a sink when $\bar{\lambda}^2 > \frac{1}{2} ( \bar{a}^2 + 6 - \bar{a}\sqrt{\bar{a}^2 + 8})$.
% Then, since all eigenvalues are always negative for  $\gamma >\frac{2}{3}$,  this point remains  a sink.

\subsection{Equilibrium Points with $\Omega\neq 0$}

There are also additional equilibrium points with $\Omega \ne 0$:

Corresponding to the flat FRW solution is
$$p_7: \quad \Psi = 0, \quad \Phi = 0, \quad \Omega=1,$$
with eigenvalues
\begin{equation}
\mu_1 = \frac{3}{2}\gamma, \quad \mu_2 = 3\gamma-2,\quad \mu_3 = \frac{3}{2}\gamma-3.
\end{equation}
Hence $p_7$ is a saddle in the entire range considered, $\frac{2}{3} < \gamma < 2$.

\subsubsection*{Equilibrium Point $p_8$:}

There is an equilibrium point, $p_8$, representing the Matter Scaling Solution \cite{coleybook}.
\begin{align}
\Phi &= \frac{1}{2\bar{\lambda}}(-\bar{a}+\sqrt{\bar{a}^{2}+9\gamma(2-\gamma)}),
\quad \Psi = \frac{3\gamma}{2\bar{\lambda}}, \nonumber \\
\Omega &= -\frac{1}{2\bar{\lambda}^{2}}(9\gamma +
{\bar{a}}^{2} -\bar{a}\sqrt{\bar{a}^{2}+9\gamma(2-\gamma)})+1.
\end{align}
The point $p_8$ has zero curvature and the deceleration parameter is given by
$$q_{p_8}=-\frac{1}{6}(3\gamma-2),$$
which is negative for $\gamma  > \frac{2}{3}$.

Linearization of the system \eqref{dPsi2} - \eqref{dOmega} about the equilibrium point $p_8$ yields the eigenvalues:
\begin{align}
\mu_\pm &= -\frac{3}{4}(2 - \gamma) \pm \frac{1}{4\bar{\lambda}}\sqrt{\mathcal{D}+\bar{a}\mathcal{E}}, \quad \mu_3 = 3\gamma-2,
\end{align}
where
\begin{align*}
\mathcal{D} & = 81\gamma^{2} \bar{\lambda}^2 - 108\gamma\bar{\lambda}^2 + 36\bar{\lambda}^2
- 324\gamma^3 -72\gamma^2 \bar{a}^2 + 648\gamma^2 + 180\gamma\bar{a}^{2} - 8\bar{a}^{2} \bar{\lambda}^{2} + 8\bar{a}^{4}\\
\mathcal{E} & = \sqrt{\bar{a}^2 + 9 - 9(1 - \gamma)^2}(36\gamma^2 - 108\gamma + 8\bar{\lambda}^2 - 8\bar{a}^2).
\end{align*}

The last eigenvalue is negative for all the values of $\gamma$ that we consider in this paper. The value of $\mu_+$ can be both positive and negative depending on the values of $\gamma$, $\bar{a}$, and $\bar{\lambda}$. But since the terms under the square root in $\mathcal{E}$ always are positive and hence $\mathcal{E}$ real, we have that the expression $\sqrt{\mathcal{D}+\bar{a}\mathcal{E}}$ is either positive or purely imaginary. The eigenvalue $\mu_-$ will therefore always have negative real part, and the point $p_8$ will always be a saddle.

\section{Notes on More General Potentials}

We also briefly take a look at more general potentials. In particular, we
consider the potential of the form \eqref{3} with a single non-zero $a_r$ ($r$ an arbitrary integer): \begin{equation}\label{general_potential}
V = V_0 e^{-\lambda \phi} + a_r \theta^r e^{(r-2)\frac{\lambda}{2} \phi}
V_0^{-\frac{(r-2)}{2}},
\end{equation}
where $a_r$ here is normalized by an appropriate power of $V_0$.  Defining normalized variables $\tilde{\Psi}$ and $\tilde{\Phi}$ as before,
$$ \tilde{\Psi} = \frac{\sqrt{3}}{\sqrt{2}}  \frac{\dot{\phi}}{\theta} \quad , \quad
\tilde{\Phi} = \sqrt{3V_0} \frac{e^{-\frac{\lambda}{2}\phi}}{\theta}, $$
the normalized Friedmann eqn. becomes
\begin{equation}
\label{Friedmann_gen_r}
1 = \tilde{\Psi}^2 + \tilde{\Phi}^2 + 2 (1-r) \tilde{a} \tilde{\Phi}^{2-r} - K,
\end{equation}
where $\tilde{a} \equiv \frac{1}{2} (3)^{\frac{r}{2}} a_r$
and we shall define $\tilde{\lambda} = \frac{1}{\sqrt{2}} \lambda$.

The normalized Raychaudhuri equation then becomes:
$$ q= 2 - (1+ r(r-1)  \tilde{a} \tilde{\Phi}^{2-r})^{-1}
(2 - 2 \tilde{\Psi}^2 + \tilde{\Phi}^2 + 2 (1 - \frac{3}{2} r) (1-r)\tilde{a}
\tilde{\Phi}^{2-r} + \sqrt{3} \tilde{\lambda}
\tilde{a} r(2-r)
\tilde{\Psi} \tilde{\Phi}^{2-r}), $$
and the evolution eqns. become
\begin{align}
 3 \tilde{\Psi}^\prime & =  -(1+r(r-1) \tilde{a} \tilde{\Phi}^{2-r})^{-1}
 \; [2-2\tilde{\Psi}^2 + \tilde{\Phi}^2+
(2-3r)(1-r)\tilde{a} \tilde{\Phi}^{2-r}  \nonumber\\
& \quad \qquad \quad
+ \sqrt{3} \tilde{\lambda} r \tilde{a}
 \; (2-r) \tilde{\Psi} \tilde{\Phi}^{2-r})] \tilde{\Psi} + \sqrt{3} \tilde{\lambda}
\tilde{\Phi}^2 - \sqrt{3} \tilde{\lambda}
\tilde{a} (r-2) \tilde{\Phi}^{2-r} \label{Psi_gen_r}\\
 3 \tilde{\Phi}^\prime & = -(1+r(r-1) \tilde{a} \tilde{\Phi}^{2-r})^{-1}
\;[\sqrt{3} \tilde{\lambda}
\tilde{\Psi} -1 - 2 \tilde{\Psi}^2 + \tilde{\Phi}^2 + 2 (1-r) \tilde{a} \tilde{\Phi}^{2-r}  \nonumber \\
& \quad \qquad \quad
+ \sqrt{3} \tilde{\lambda} \tilde{a}r  \tilde{\Psi} \tilde{\Phi}^{2-r}] \tilde{\Phi}. \label{Klein-Gordon}
\end{align}
For $r=1$ these equations reduce to the equations studied earlier.

Let us consider equilibrium points with zero curvature. If $K = 0$, we obtain from \eqref{Friedmann_gen_r}
\begin{equation}
\label{curvature}
\tilde{\Psi}^2 = 1 - \tilde{\Phi}^2 - 2 (1-r)\tilde{a} \tilde{\Phi}^{2-r}.
\end{equation}
Setting the right-hand sides of \eqref{Klein-Gordon},\eqref{Psi_gen_r} to zero to obtain the
equilibrium values $\tilde{\Phi},\ \tilde{\Psi}$ and using \eqref{curvature}, assuming
$\tilde{\Phi} \neq 0$, we find after some algebra that $\tilde{\Phi}$
satisfies the polynomial eqn.:
\begin{equation}
\label{EqnA1}
\tilde{\Phi}^2 =  1 - 2 (1-r)\tilde{a} \tilde{\Phi}^{2-r}
 -\frac{\tilde{\lambda}^2}{3} (1+r \tilde{a} \tilde{\Phi}^{2-r})^2,
\end{equation}
whence we obtain
\begin{equation}
\label{EqnA2}
\tilde{\Psi} = \frac{\tilde{\lambda}}{\sqrt{3}} (1 + r\tilde{a}
\tilde{\Phi}^{2-r}).
\end{equation}
It can easily be checked that these values for $\tilde{\Phi}$, $\tilde{\Psi}$ do indeed satisfy the zero-curvature condition \eqref{curvature} and are equilibrium values of the system \eqref{Klein-Gordon}.  For vanishing $\tilde{a}$, we obtain the usual zero-curvature inflationary power-law equilibrium solution $P$ in standard exponential potential cosmology $(\tilde{\Phi}= \sqrt{1- \frac{1}{3} \tilde{\lambda}^2}, \tilde{\Psi} = \frac{\tilde{\lambda}} {\sqrt{3}}$), which is asymptotically stable to the future. Linearization of the system \eqref{Psi_gen_r}, \eqref{Klein-Gordon} is more complicated with a general power $r$, but an approximate linearization for small $\tilde{a}$ around the zero-curvature solution is possible\footnote{A work in progress by M. Stevens.}.

\section{Discussion}

% Summary.
We have analyzed the dynamical evolution and stability of inflationary solutions of homogeneous and isotropic Einstein-Aether cosmologies containing a scalar field, under the assumption that the scalar field interacts with itself and the aether through a potential given by \eqref{33}. The potential is of the general form \eqref{3} proposed in \cite{Barrow} and more general than the examples explicitly studied therein.

% first potential
We find that the scalar field-aether interaction term only slightly affects the stable solution of the system for small values of the normalized scalar field self interaction coupling $\bar{\lambda}$, even when the normalized scalar field-eather coupling $\bar{a}$ is large. When large scalar field self interactions are considered there is a qualitative change in the future stable equilibrium of the system compared to the ordinary exponential potential case. For negative values of the coupling $\bar{a}$ there exists a region (region II, Fig. \ref{sinkplots}) where the stable equilibrium is spatially flat but non-inflationary, a situation that in the normal case requires fine tuning of $\bar{\lambda}$. Even larger scalar field self interactions destabilizes the flat equilibrium and drives the state towards a spatially curved equilibrium, like in the normal exponential potential case.

% influence from matter
When we also include a matter source term in the form of a perfect fluid with linear equation of state that does not couple directly to the other fields, we find equilibrium states corresponding to the usual FRW model and matter scaling solution. The precise value of the normalized field variables of the latter equilibrium depend on the coupling $\bar{a}$, but the qualitative properties like its curvature and deceleration parameter do not. The equilibrium point is a saddle when curvature perturbations are considered, like in the normal case, but even within the flat models it can remain a saddle for some values of $\bar{\lambda}$ and $\bar{a}$, unlike the case with an ordinary exponential potential where the matter scaling solution is a late time attractor. The matter source term do not qualitatively alter the stability of the sinks found earlier, they all obtain a stable manifold of one dimension larger than previously and thus remain sinks.

% general potential
Also potentials with a coupling between the exponential part and the expansion of general order are susceptible to a dynamical systems analysis in the scale invariant variables. The problem of finding a scale invariant solution with zero curvature can be reduced to solving a polynomial equation, and the solutions all reduce to the standard inflationary power-law solution when the coupling becomes arbitrarily small.

\subsubsection*{Acknowledgments:}

We thank M. Stevens for helpful insights on the general $r$ potential. PS also thanks AAC and the department of mathematics at Dalhousie University for their kind hospitality. AAC is supported by grants from the Natural Sciences and Engineering Research Council of Canada. BA would also like to thank the Government of Saudi Arabia for financial support.

\end{document}